# A Frame for Communication Control

*Aernout Schmidt[1] & Kunbei Zhang[2]*

## Abstract

1. We are experiencing the rise of ChatGPT-like systems or LLMs in political turbulent times.
2. We assume the need to regulate their use because of their bubble-shaping and polarizing potential.
3. To regulate, we need a language that allows interests and compromises to be discussed.
4. In this context, we can think of such a shared language as a *jargon,* a specialized vocabulary for law-making.
5. To the extent such a jargon exists, it is now being corrupted by LLMs, see point 2.
6. Points 2–5 appear paradoxical.
7. The issue includes persistent communication failures, between disciplines that cannot translate their technical vocabulary into accessible terms, and between political movements that operate in incompatible worldviews.
8. We show that a frame integrating four specialist languages, those of governance, economy, community and science, is able to address these failures case-wise, which we consider helpful.
9. However, for reasons noted in Point 3, we cannot deploy the frame to create the more generic jargon on our own.
10. We conclude that our frame provides the knowledge to design and apply RAG-LLM architectures for researching its jargon generating potential in a future project.
11. We show its feasibility in the appendix.

10,216 words, 3 tables.

**Keywords**: Failing governance debate, Diplomatic Framing, AI-supported dialogue-failure analysis, Cross-institutional NLP, Everyday Language AI support, Retrieval-Augmented Generation (RAG).

**Subject Categories**: Managing AI, Managing with AI, AI supporting governance debates, Free Speech Framing, Retrieval-Augmented Generation (RAG) Design, Dioplomatic Constraints.

## 1. The Frame

We're living through turbulent times in 2025. Stalled debates between rule-makers are a symptom. We can't offer a solution, especially not when for some parties a solution isn't welcome anyway. What we can do from a scientific position is propose and examine an instrument that can help analyze positions in debates neutrally. And thereby search for a jargon that doesn't easily get stuck. That instrument is not everyday language, not an idiom (or professional language), not a dialect (or geopolitical regional language), but a frame.

By frame we mean, following Rein & Schön [1] an interpretive framework through which conversation participants select what is relevant, define how something works, and evaluate what is good or bad, desirable or undesirable. Framing has long been established in political debate, though it has gained a bad reputation among those who value academic neutrality. Yet since framing is unavoidable in political discourse, we argue that identifying and studying a broadly acceptable, neutral frame is an important task for the humanities.

---

[1] Professor emeritus at Leiden University, Center for Law and Digital Technologies, The Netherlands. ORCID No.: 0000-0003-2594-5767. Email: schmidtahj@umail.leidenuniv.nl

[2] Associate professor at Chongqing Technology and Business University, Faculty of Law and Sociology. ORCID No. 0009-0000-9798-1615. Email: zhangkunbei@ctbu.edu.cn

## 1.1  The Accountability Gap, the Research Question

We draw on the accountability gap, first highlighted by the Freenet project in 1999. This gap has only been amplified by the AI revolution of 2023. Together these lead to an an increasingly irreconcilable conflict—one that underscores the need for a frame like J4CC to analyze incompatible debate positions. We have many conflicts of the kind and their number is increasing, such as over geopolitical borders, immigration, emergency aid, climate change.

### Freenet (1999)

The Freenet project offers a secure platform for information exchange. Clarke [2] modeled four core principles. *Decentralization*, no central server; *anonymity*, nodes don't know the complete data path; *data persistence*, popular content is preserved while rarely accessed content disappears; *censorship resistance*, no central authority can remove content. When Freenet started, it was a working peer-to-peer system that strongly appealed to activists, dissidents, and privacy advocates.

But within a few months it also attracted criminals—those spreading child pornography, illegal marketplaces, and extremist propaganda (personal observation, end of 2000). Precisely its strength—the anonymity of who inserted or retrieved a file—proved a significant challenge for managing information flows. Without a central server to shut down and without built-in oversight mechanisms, institutions struggled to assign responsibility.

Thus Freenet demonstrated that there is a demand for communication control (CC), even in that part of the world that cherishes communication freedom (CF) as a human right. The fundamental problem is that those responsible for message content are secret (as argued, for example, by the U.S. Department of Justice in its report to Congress [3].

### The AI Revolution (2023)

This problem is now magnified by the widespread deployment of large language models or LLMs, which inject AI results into what search engines and other social media deliver. Since late 2024, also unsolicited (personal observation). Today's LLMs thus obscure the traceability of information sources, lifting the digital accountability gap to a global scale. The accountability gap thus leads to what feels like a rapidly expanding legitimacy crisis.

We illustrate its urgency through a phenomenon that so far seems to escape regulation. Government and AI appear to be joining forces increasingly. This leads to every jurisdiction being confronted with the question of how political, economic, cultural, and knowledge-based power relations develop under AI use becoming ubiquitous. Since March 2023, Western LLMs, for example from OpenAI, Anthropic, and Google, apply forms of censorship and content filtering. This has led to booming business. Scale AI, for instance, a 2016 startup that preprocesses or labels training data by human hands, had 49% of its shares purchased by Meta for $14.3 billion in June 2025 [4]. With this, Meta bought labeling power. This matters to us not only because labeling of training data functionally equals AI framing. It also matters because this way of working can become problematic if how labeling power is deployed remains secret and/or is used to censor, or to profile users while further deepening their dependency on social media platforms.

Naturally, thinking about how the digital accountability gap works requires us to take a position. We need language for that. And at the moment there's no established jargon yet. J4CC is intended to grow one.

## A Simplified Example: East and West

Think, for example, of the CC/CF contrast between China and the US, circa 2023. Imagine the challenge when service providers want to comply with the rules. American internet culture has traditionally emphasized openness and decentralization. China has seen cyberspace from the beginning as a territorial thing, complete with borders, institutions, and enforcement systems. Technology that enables anonymity is considered a source of social instability. China has therefore developed multilayered censorship and surveillance systems that block or filter tools like Freenet. It's collectively known as the Great Firewall. This is not only multilayered but stratified—like geological layers that have folded and compressed together, creating a complex system where technical, legal, and behavioral controls intersect and reinforce each other. This Firewall includes deep packet inspection, IP blocking of known services (e.g., Tor, Google, Wikipedia), and legal regulation as in the Cybersecurity Law (2017) and Data Security Law (2021). These require identity registration, data localization, and state access to encrypted content. They're supplemented by behavioral influence that combines AI with human oversight and integrates with social credit systems to discourage anonymous or subversive activity. A Freenet-like service spreading illegal content would, whether separatist rhetoric, historical revisionism, or liberal-democratic propaganda, provoke immediate enforcement ranging from digital surveillance to criminal prosecution. The governmental axiom is that technological anonymity makes CC enforcement impossible and raises the risk of social unrest.

The traditional Western approach (for example, codified in the US in CDA § 230, from 1996) relies largely on market mechanisms for communication regulation. The basic logic is: "You can't enforce CF just a little bit. Whoever or whatever censors can abuse that power to suppress truth." Compared to this, China's governance model is proactive. The AI governance model (crystallized in 2023) shows a fundamentally different approach there: by requiring ideological alignment, prohibiting unauthorized historical interpretations, and directly integrating politically sensitive filters into domestic LLMs like Baidu's Ernie Bot and SenseTime's generative systems.

Different jurisdictions thus respond in radically different ways, showing divergent perspectives on how information flows should be regulated. WeChat, an app that complies with China's traceability requirements, violates the US privacy rules that WhatsApp claims to follow.[3] We suspect the Chinese Politburo and associated services wouldn't think of using WhatsApp for the same reason (if its use weren't banned in China[4]).

## The Question

There are more than two jurisdictions in the world, and they can differ radically in how they respond. For service providers and users, it's nearly impossible to identify the regimes that apply to their communication channels. What's needed is a jargon to describe the differences between regimes. Challenges such as those Moss [6] tackle through scenario development and offering structured storylines for an uncertain future. Their focus is on regulation that faces climate change risks. We want jargon to face language risks. Risks such as those raised by the ubiquitous deployment of potentially polarizing and addictive, LLM-injected communication channels (Pazzaglia et al. [6]). In times of geopolitical unrest, moreover (the emerging Russia-Ukraine and Israel-Iran wars, June 2025). We summarize our problem statement as follows:

> *Can we meaningfully analyze domestic and cross-border debates about communication control and the accountability gap by developing a public jargon with a frame tailored to this purpose (J4CC)?*

---

[3] There's different thinking about this too. According to an Axios report [5] from June 23, the US House Office of Cybersecurity labeled WhatsApp as very high-risk in an email: "...high-risk to users due to the lack of transparency in how it protects user data, absence of stored data encryption, and potential security risks involved with its use" and: "House staff are NOT allowed to download or keep the WhatsApp application on any House device, including any mobile, desktop, or web browser versions of its products" and: "If you have a WhatsApp application on your House-managed device, you will be contacted to remove it." But here national security is the issue rather than privacy.

[4] For that reason, we (the authors) use WeChat to communicate: AS from the Netherlands, KZ from China. WeChat is (still?) legally operational in these jurisdictions.

Debate participants are authorities as well as service providers as well as (collectives of) believers, association members, entrepreneurs, farmers, workers, scholars, artists, citizens, and consumers. Every jurisdiction has these blood types.

A difficulty is that there is no math model available when we must choose a rule. When we must simultaneously weigh the interests of public order, the economy, culture, and science, we cannot trust a formula to do the work but are dependent on thinking and deliberating ourselves in everyday language.

We're not choosing between goods or services—we're choosing rules. This requires weighing four different kinds of value at once, like: domain control, economic benefit, social solidarity, and factual accuracy. Because three of those four currencies cannot be expressed in money, choices must be made.[5] The difficulty is naturally that we must weigh different interests for which we have no common scale.

Wait—are we repeating ourselves? No. We need to flip how we're looking at the problem. The difficulty is that we emotionally want a common measuring stick. But we shouldn't want a measuring stick for interests that can't be measured that way. What we're investigating is whether AI models can help us navigate this fundamental incommensurability. After all LLMs navigate the fundamental incommensurabilities that would otherwise prohibit participation in natural language debate. And its abilities to do so emerge from word frequency statistics drawn from natural language training material.

## 1.2 Frames as Jargon Filters

Rules must emerge from practice and not be proposed or imposed from outside or from above. Our normative position follows Conklin's [10] view of the Hart–Fuller debate: rules don't arise before practice (contra, Hart [11]) but emerge from it (Fuller [12]). Ergo, rules must be forged in debate between stakeholders, in a shared language. This may seem hopelessly idealistic when so much public debate has collapsed into online vitriol and personal attack. But the urgency of the debate cannot be denied—particularly now, as jurisdictions and institutions worldwide rush to regulate AI.[6]

If rules must be forged in debate between stakeholders and in a shared language, this brings us to a core question: Isn't there already a language that allows diverse stakeholders to deliberate about not-yet-formulated rules for not-yet-defined risks caused by LLMs?

We believe so: *everyday language.*

Not the specialized jargon of law or computer science, but the everyday language in which AI risks are already being discussed worldwide—in China, the US, Europe, and the scientific community, for example. The discussions take place daily, in talk shows, newsrooms, social media and, occasionally, in parliamentary debate. These discussions are certainly also conducted prominently in the spaces where LLM-related services are designed and developed by people.

---

[5] It is precisely this problem, that choice also entails costs whose values are not or cannot be expressed in monetary terms, that Coase [8] raises. This occasionally leads to misunderstandings when the value of welfare is conceptually limited to what can be expressed in monetary terms (e.g., Kaplow & Shavell [9]).

[6] For example: the Bletchley Declaration of the UK AI Safety Summit of November 2023; China's AI Governance Framework of 2023; the EU AI Act adopted on June 13, 2024; the California Report on Frontier AI Policy of June 13, 2025 – but also, indirectly, through the rise of labeling policies by services with market power such as AI Scale.

But if the language exists, *what's the problem*?

The problem is that everyday language has proven unsuitable for the job. As recent reactions to conflicts over territorial sovereignty, climate change, migration, genocide, pandemics, and AI risks have shown, everyday language as such is insufficient because it facilitates the emergence of taboos, impasses, power plays, and shouting matches whose use distorts or blocks communication. We need a more productive jargon.

But *we* can't make that—it must emerge in practice. What we can do is propose a frame that can help steer the formation of that jargon in that manner. Functionally closely related, we think, to what emerged through long practice in the diplomatic world and was commonly used until recently. We're looking for a frame that, with the help of AI, produces such a jargon while being used.

Here follows our proposal.

## 1.3 The J4CC Frame

We assume that at least four forces simultaneously structure CC practices. Conventionally, analysis is often limited to two zones: East (Power/Sovereignty) and West (Market/Capital). But we see four (Power, Capital, Morality, and Knowledge). We first widen the view from two to three: Power-oriented zones (e.g., China), Capital-oriented zones (e.g., the US), and Morality-oriented zones (e.g., the EU, where the rule of law is in focus). They form three ecologies; we summarize them in Table 1.

Table 1: Three J4CC Zones Linked with Eight Important Characteristics

| Core Attractor | Power (China) | Capital (US) | Morality (EU) |
| --- | --- | --- | --- |
| Core Currency | Domain control | Economic benefit | Social solidarity |
| Core Goal | Control, Security | Profit | Human Dignity |
| AI Infrastructure | State-owned | Private sector | Hybrid |
| Language Use | Ideologically aligned | Profiled & labeled | Free, human right |
| Who Builds AI? | Tech (State oversight) | Big Tech (mil.-industrial complex) | Open Source |
| Mechanism | PRC democracy | US democracy | EU democracy |
| Primary Risk | Power monopolies | Market monopolies | Loss of support |

While Table 1 provides a compact overview of forces that attract CC-related propositions, something is missing: knowledge. Not politicized knowledge, but operational, propositional knowledge—the kind that science and projects like Wikipedia embrace. The corresponding attractor (pulling force) hasn't been mentioned yet. So there's a missing zone where truth and neutrality are the primary issues. Recognizing that science cannot flourish where Power, Capital and/or Morality call the shots, we introduce a fourth zone: Propositional Knowledge – like that which provides a foundation for Popper's [13] argument and like that which underlies what are called *facts* and illustrate their attractive force through the emotions that, for many, drive and emerge from *fact checking*. This fourth zone requires its own attractor: Knowledge. We therefore expand our frame with Table 2, gaining a four-zone model alongside Power (e.g. China), Capital (e.g. US), Morality (e.g. EU), now also Knowledge (e.g. in academia).

Table 2: The Fourth Zone

| | |
|---|---|
| **Core Attractor** | Knowledge (Academia) |
| **Core Currency** | Propositional truth |
| **Core Goal** | Describe/Predict |
| **AI Infrastructure** | ArXiv, PyTorch etc. |
| **Language Use** | Paradigmatic, Propositional |
| **Who Builds AI?** | Students, Start-ups |
| **Mechanism** | Peer review |
| **Primary Risk** | Loss of support |

The complete frame introduces a platform for discussing sociopolitical issues within and between zones. Each zone has a distinctive CC profile, with all four attractors present in varying weights. Our goal is to find adequate linguistic means, a jargon for describing interests and negotiating between stakeholders from these ecologies. The frame enables us to map both simple and very complex information-transfer ecologies by identifying which forces operate where, with which institutional connections, and under which constraints (see Section 2). Everywhere in an institution's force field, four attractors are felt simultaneously; how strongly depends on their size and proximity. This is the idea of an institutional four-attractor force field. That this field is stratified results from recognizing that institutions can combine and overlap and span aggregation levels.

## 1.4  Zones, Fields, and the Four Attractors

We spoke of zones and attractors as if that went without saying. And perhaps you've also formed an image of them, because we named zones (China, USA, the EU and academia) and corresponding goals (order/security, profit/capital, human dignity, and descriptive/predictive potential). We still want to elaborate a bit.

### The Zones (domain, profit, solidarity, and truth)

It is indeed our intention to see zones as areas where the mentioned four forces simultaneously exert their attractions. We see these as magnet-like things that, depending on their position and strength, influence every place in the field where rules are discussed. A zone is then a field with four fixed attractors, one of which is the largest and therefore dominant in the region. In the US, money/capital is the largest attractor, and in China it's order/security. Depending on position in the zone, there's a different configuration of attractor forces. We could express this in a list.[7] For example: [os,pc,hd,dp] for the attractors order/security, profit/capital, human dignity, and propositional truth respectively; suppose for each a three-value scale is available, then [2,3,1,1] could be a value profile for an argument (or debate participant, or an entire jurisdiction). The list indicates that this person or thing is moderately attracted by order/security, strongly by profit, and barely by human dignity or propositional truth.

The first three zones (os,pc,hd) are geopolitically recognizable; the fourth (dp) is of a different nature. It overlaps all three other zones.

### Fields

We assume that fields represent jurisdictions, that is, areas for which rules can be made and enforced. Each field has its own attractors and its own profile. We therefore assume that jurisdictions can be stacked and can overlap each other, including cross-border. Our consciousness can thus be a field, just like a chess federation and just like the US.

---
[7]  Technically: a vector (sometimes we will use that term.)

## The Four Attractors

*Power (Control, security – domain/order).* Power is a coercive force aimed at control and security, hierarchically organized, that guards the geopolitical domain and delivers order.

*Capital (Profit, capital – market/opportunities).* Capital is force aimed at economic profit and prosperity, organized so that the price mechanism, the principle of comparative advantage, and the level playing field make their contribution. This means non-hierarchical organization that promotes opportunities, competition and innovation.

*Morality (Human dignity – solidarity).* Morality is the force that binds and blinds groups, seeking a dignified identity. Haidt [14] distinguishes personal, intuitive fields where feelings of danger, equality, authority, solidarity, and purity yield a profile. This profile emerges partly based on 'nature' and partly on 'nurture.'

*Knowledge (propositional truth: description/prediction – method/neutrality).* Popper [13] rejected verification as a criterion of science. Instead, he proposed that scientific statements must be falsifiable. In Popper's view, we can only speak of knowledge when it is propositional, public, and falsifiable. This conception has held up in our time in the form of fact-checking. We choose this form of knowledge as attractor because the oft-heard "scientific knowledge is just another opinion" overlooks the methodological requirements of scientific practice—requirements that opinions don't meet.

## J4CC as Jargon Filter

At this stage, a claim to precision would be counterproductive. We use concepts with high family-resemblance potential and thus still need much room for empirical filling and interpretation. We avoid philosophical debate but acknowledge that we drew our inspiration here from two classic sources: Wittgenstein's [15] family resemblance and Nietzsche's [16] truth as lies in a non-moral sense. And from a recent source, Hodgson's [17] informal institutions. Still—we're almost far enough to move toward case studies to show that the frame can be used to sketch and compare divergent profiles and thereby develop the beginning of a jargon. Thus becoming a 'jargon filter.'

But that term alone can evoke anti-feelings that are fed by different constitutional convictions. Discussions about regulating CF (communication freedom) and CC (communication control) get stuck easily, after all.

In short, we're dealing with a hitherto unsolvable question.

As we haven't yet shown our vision of how AI could help we give a sketch of how we think J4CC can work as a jargon filter. What we need is:

- An LLM (like ChatGPT, Claude,...)
- A library of J4CC-filtered (labeled and vectorized) specialized natural-language knowledge like negotiation records as part of the J4CC application
- A prompt (description of the regulatory question and the starting positions of a party)

And what we expect is:

- A J4CC-generated prompt that is submitted to the LLM
- The LLM's contribution

After which an assessment follows that is added to the library of filtered specialized knowledge along with the LLM contribution.

## 1.5 Theoretic Foundation

We note that our approach is inherently multidisciplinary and requires further theoretical clarification here and there. Creating a frame for AI governance debates requires drawing from (i) legal and political theory: e.g., Confucius [18], Fuller [19], Bobbitt [20]; (ii) economic theory: e.g., Ronald Coase [21], Milton Friedman [22]; (iii) anthropology: e.g., Montesquieu [23], Pierre Bourdieu [24], Mary Douglas [25], Jonathan Haidt [14]; (iv) philosophy of science: e.g., Karl Popper [13], Kuhn [26], Philip Anderson [27] [28], Ian Goodfellow [29]. This disciplinary breadth isn't academic sprawl but necessity (see also our theoy-based take on multidisciplinary research below). Each field (we limit our references to what we consider a minimum) contributes essential insights. And each field cherishes its own meaning-making methods: legal theory explains institutionally how legitimate norms arise and hermeneutically what they mean; economic theory shows, through statistical modeling combined with empirical observation, how the price mechanism works and which conditions disrupt it; anthropology shows, often through participant observation, how collectives and collective norms emerge and function; science shows, often through mathematical modeling and empirical observation (but recently also through pragmatic study of AI application behavior), how physical systems and computer models and languages "work."

The challenge is to filter and recombine these disciplinary vocabularies into a jargon in everyday language that can support debate (including among citizens) about rules for AI use, between participants from different jurisdictions and value systems.

However.

In practice, this will inevitably produce new forms of technical language, because communities naturally develop specialized languages that function as framing machines (Lakoff [30]; Entman [31]). And professional languages lead to tunnel visions that are often at odds with broader political debate, thereby substantively limiting communication again (Rawat [32], Kuhn, [26], Habermas [33]).

In short, we're dealing with a hitherto unsolved question. We must acknowledge that our goal seems impossible: developing a frame capable of transforming both everyday language and professional languages into jargon suitable for political debates about communication controls (CC) and communication freedom (CF). Such jargon should maintain accessibility for both specialists and laypeople across diverse legal and cultural jurisdictions and institutions—a challenge that lives at the intersection of science communication theory (Fischhoff [34]) and comparative legal linguistics (Tiersma [35]). There too, the question remains largely unanswered. Consequently, our question is unanswered.

However.

Because it has recently become possible and feasible to deploy AI and Lewis et al.'s [] RAG-enhanced transformer-based systems[8] to further investigate this gap and ultimately bridge it pragmatically (see Goodfellow's profile in Section 2), we're working on finding an AI-supported path to such a solution.

To explain why we chose some of the literature mentioned, we give a few of the deliberations, at the boundary of science and the humanities, that we've built upon.

### Coase - Institutions

In his search for the origin of firms, Coase [21] thought about the conditions under which it makes sense for an entrepreneur to hire an employee rather than buy their intended output on the market. His analysis rests on weighing different types of costs and concludes that firms emerge as soon as hiring employees costs less money. (The excitement about this arose because it was then assumed that the market would always be more efficient, overlooking that using the market brings costs too.) Coase's [8] analysis led to institutional economics, a branch of economic science where the original attention to firms has broadened to attention to institutions, so much so that the idea of rules has become central in answering the question of whether there is economically an institution. Those institutions have the rules. Institutions are (not necessarily nationally defined) jurisdictions. They are domains where rules apply. Like bridge tables during a bridge drive. We use that notion of institutions.

---

[8] A few technical terms referring to the architecture of large AI language models that have not yet entered everyday language.

## V. Ostrom - Polycentricity

On every J4CC field, four attractions are felt. We know them by now, the attractive forces of Power, Capital, Morality, and Knowledge. But they come from multiple sources. Not only from sources on their own field, but also from sources on neighboring fields. If every field has its own four force sources, every actor feels, for every argument, the assembly of forces working on it. Every institution thus generates its own profiles, shaped by a distinctive configuration of four attractive forces. Because domains can border each other and also overlap each other, this means the force field is stratified. Like—according to Stephan et al. [36]—the institutional characteristics in V. Ostrom's polycentric regulation. These correspond to the forces in our frame. We quote: "According to V. Ostrom, the concept of polycentricity encompasses economic markets, legal orders, scientific disciplines, and multicultural societies." We have, unwittingly, chosen V. Ostrom's governance forces from 1960 for our frame. And we use his approach to polycentricity for our image of stratified force fields.

## Anderson - The Generative Potential of Complexity

We borrow from the broad, primarily natural-science oriented literature on complex systems a notion we call the generative potential of complexity, which we recognize in the behavior of large language models. We need a theoretical bridge here. For conceptual orientation, we turn to Philip Anderson's "More is Different" [27] and "More and Different" [28]. These challenge the idea of unified science and instead propose a stratified epistemology. Anderson argued that complex systems exhibit broken symmetry, where small changes at lower levels can produce qualitatively new behaviors at higher ones (e.g., superconductivity where the symmetry of some crystals is broken, 'contaminated'). He doesn't claim that lower-level analysis is impossible, but that higher-level patterns are autonomous and require new concepts. They are emergent patterns, invisible from the analytical perspective. Like an incidental idiomatic surprise in a text can revive the attention of a dogmatically stifled reader. We're actually counting on the generative capacity of transformer models[9] to help us find patterns in everyday language that form the jargon we're looking for.

## Multidisciplinary research

The J4CC frame's vector-based approach gains theoretical depth by incorporating established sub-models from our contributing disciplines as concrete handles for future[10] environment and agent modeling. Bobbitt's [20] six constitutional interpretation methods (historical, textual, structural, doctrinal, ethical, prudential) provide granular coding mechanisms for legal-institutional environments, while Fuller's [19] eight principles of legality (generality, promulgation, non-retroactivity, clarity, non-contradiction, possibility of compliance, constancy, congruence) offer precise institutional quality metrics. Douglas's [25] group-grid cultural theory yields four distinct organizational cultures (hierarchist, individualist, egalitarian, fatalist) that map directly onto institutional environment vectors, capturing how power relations and social boundaries shape governance contexts. Similarly, North and Hodgson's [17] institutional analysis provides operational criteria for distinguishing formal from informal institutions, while Haidt's [12] moral foundations theory contributes five emotional-cognitive dimensions (care/harm, fairness/cheating, loyalty/betrayal, authority/subversion, sanctity/degradation) that enable nuanced profiling of moral attractors within specific cultural contexts.

---

[9] Another technical term referring to the architecture of large AI language models that has not yet entered everyday language.

[10] This addresses the disciplinary complexity bridging our two-stage approach: establishing J4CC as an analytical **frame** (this article's focus) and developing J4CC as a working **jargon** through RAG-based AI implementation (our planned follow-up project, introduced in the Appendix). The multidisciplinary integration sketched here provides the theoretical foundation necessary for both stages.

These established theoretical sub-models function as building blocks for constructing detailed vector representations of agents and environments within J4CC's four-force field. Rather than relying on impressionistic coding, the frame can leverage decades of empirical research validating these dimensional approaches. When implemented in LLM-supported applications, these sub-models provide structured pathways for translating qualitative governance phenomena into quantitative vector representations, enabling systematic analysis of cross-institutional dialogue patterns while maintaining theoretical rigor across disciplinary boundaries. This approach transforms J4CC from a conceptual frame into a practically implementable analytical system capable of supporting real-world governance applications.

Here we conclude the introduction to the frame and move on to a few case studies.

## 2. Empirical Exploration

We summarize our research question as follows: Can we effectively analyze stakeholder debates about communication control (CC) and the accountability gap by developing a dedicated jargon with J4CC? After establishing our frame as an institutionally stratified, four-attractor force field suitable for tackling with AI tools, we give an impression of how it can be used as shorthand for profiling and coding. We start simply: How might some of the important thinkers from diverse intellectual traditions mentioned above position themselves in relation to J4CC's four forces—Power, Market, Morality, and Method? The case study approach is for discovering the frame's prima facie expressive qualities, for profiling individual situations and agents, and their environments.

Profiling thinkers with J4CC enables us to see how the frame captures both established traditions and ongoing conflicts in a language that remains tolerable across political dividing lines. J4CC works like GPS—it doesn't eliminate the journey, but it does make the territory visible.

### Coding

We use compact six-value vectors or lists. For each thinker we generate two: one that profiles their personal position based on their publications and one that profiles their environment based on their institutional-political context of that time. For source material we used Wikipedia. Each list follows the format: [ID,Type,os,pc,hd,dp].

For example, [Confucius,env,1,2,3,1]—profiles the environment (community, jurisdiction) in which Confucius lived with four values for the attractors in his environment: low influencing by order/security, moderate influencing by money/capital, strong infuencing by convictions on human dignity, and low attraction by propositional knowledge.

A second vector—[Confucius,pub,3,1,3,2] describes Confucius's profile based on his publications: strong emphasis on order/security, low emphasis on money/capital, strong emphasis on human dignity and medium emphasis on propositional truth.

Note that where differences between the two vectors occur, one might feel inclined to look for individual motivations and for zonal (community) risks.

The case study method is for discovering the frame's *prima facie* expressive qualities, for profiling individual agents and their environments. This approach may seem informal and imprecise. We agree, but are not impressed. We aim to capture the essence of political debate positions here—what agents care about, and what their environments care about, not whether what is coded is objectively true. Moreover, when we would train a model, an option we discuss in the Appendix, we could begin—as many language models do—with random parameter values that only later acquire more precise meaning through learning from enormous amounts of text.

## 2.1 Thinkers

We claimed that our frame is instrumental. So it should at least be able to politically profile diverse agents in diverse jurisdictions. In this subsection we investigate how the frame can be used in this way. We ask: How might important thinkers from diverse intellectual traditions position themselves in relation to J4CC's four forces, in their respective institutional contexts at the time? An how can we profile their ambitions, based on what they published? We chose five thinkers and linked them to the attractor we expected to be the most important for them: Confucius [os], Milton Friedman [pc], Montesquieu [hd] and Karl Popper [dp]. We added Ian Goodfellow [dp] who is a front runner in AI and whom we guessed (unwarranted as we will see) to side with Popper.

Here we go.

**Confucius**

```
env,1,2,1,1
pub,3,2,3,2
```

Confucius (551–479 BCE) lived during China's Spring and Autumn period (770–476 BCE), a turbulent era marked by political fragmentation, shifting alliances, and frequent warfare among numerous small states nominally under the declining authority of the Zhou dynasty. Confucius lived in a politically fragmented era, where centralized authority eroded and smaller states vied aggressively for dominance. His philosophical outlook and political activism responded directly to the chaos and disintegration of traditional Zhou order, making his institutional environment both challenging and stimulating for the development of his influential moral-political philosophy.

Confucius' [18] morality emphasizes hierarchy and Power primarily because he believed social order, peace, and individual moral development depended on clearly defined and morally maintained roles within a stable political and familial structure. Several interconnected reasons help explain this emphasis. Confucius' morality places high regard on hierarchy and Power because he saw them as essential institutional frameworks enabling individuals and society to realize moral potential, maintain order, and prevent social disintegration. This outlook is not authoritarian for its own sake, but rather grounded in the belief that well-structured hierarchies and moral sovereignty yield a stable, virtuous, and harmonious society.

> *Confucius would feel at home near a strong Power and a strong Morality pole, both close together. His preferences may well tally with his disgust for the chaotic character of public power and moral arrangements in his time.*

**Milton Friedman**

```
env,3,2,3,2
pub,1,3,1,2
```

Milton Friedman (1912–2006) Lived through the Great Depression, World War II, the Cold War, and the rise and fall of Keynesianism. His political circumstances and institutional environment deeply shaped his defense of economic freedom and limited government, and his moral philosophy placed economy and economic welfare at the heart of individual liberty and responsibility. His political circumstances emerged from depression-era regulation being followed by Cold War ideological combat—which nudged a philosophy that views free markets not just as efficient, but as morally just. During the Cold War, the global battle between capitalism and communism framed economic debates as existential.

Friedman's [22] work was a forceful defense of market liberalism against central planning. He argued that political freedom depends on economic freedom, linking his economic theories to broader moral and political stakes. Milton Friedman sharpened this logic into a libertarian principle: minimal interference yields optimal outcomes. He argued that free competition, whether among firms or ideas, is the best validator of truth. In this zone, information is judged not by correspondence with fact, but by performance. Truth sells, or it doesn't. Validation comes through market survival, not epistemic consensus. Friedman's economic philosophy is inseparable from his moral vision —a belief that free markets are morally superior because they protect liberty, promote responsibility, and reduce coercion.

*We guess that Friedman would prefer to live near the Market attractor pole, with Knowledge in the background and wishing the others located at even furthher distance. He was educated in a time where Power and Morality (independent from Market) dominated with less focus on Market and Knowledge.*

## Montesquieu

```
env,3,2,2,2
pub,2,3,3,2
```

Montesquieu (1689–1755), one of the key Enlightenment political philosophers, lived in a time of absolutist monarchies, colonial expansion, and growing debates about constitutionalism, commerce, and liberty. The French monarchy weakened traditional institutions like the *parlements* (regional law courts) and nobility, replacing them with centralized royal bureaucracy. Montesquieu's thinking was deeply shaped by the early modern colonial world and its reports on non-European societies.

He used comparative political anthropology to argue that laws must vary according to geography, climate, religion, economy, and culture—a relativistic turn rare in Enlightenment thought. Montesquieu's [23] central idea is that morality must be embedded in institutions appropriate to the "spirit of the laws" of a given people. He rejected universal, one-size-fits-all legal or moral codes. Laws reflect the spirit (esprit) of a people—a synthesis of history, geography, religion, economic conditions, customs, and character. Therefore, morality must be institutionally adapted: what is just in one society may be unjust or absurd in another. Montesquieu believed liberty depended not on abstract rights, but on a structural balance among institutions.

*We think that Montesquieu would prefer to live close to the Morality and Market attractor poles, yet not out of reach of the others. In his time Power towered above the other forces.*

## Karl Popper,

```
env,3,1,3,1
pub,1,2,2,3
```

Karl Popper (1902–1994) lived through the intellectual upheavals of the early 20th century, including the collapse of liberal empires, the rise of totalitarian ideologies, and the transformation of science and philosophy. Born in Vienna, Popper came of age during the aftermath of World War I, as liberal monarchies gave way to radical ideologies—communism, fascism, and Nazism. He was deeply affected by the rise of authoritarian politics, especially the Nazification of Austria and later, the suppression of dissent in the Soviet Union. As a Jew and anti-totalitarian, Popper fled Austria in 1937, eventually settling in New Zealand, and later becoming a professor at the London School of Economics. In Britain, he was embraced by liberal intellectuals defending open societies against both fascism and Soviet communism.

Popper's [13] claims that truth, especially in propositional form, is essential to protecting both science and democracy from dogma, propaganda, and authoritarian closure. His commitment to propositional truth reflects both philosophical rigor and a deep moral-political concern with safeguarding intellectual freedom. Popper famously rejected verification as a criterion of science. Instead, he proposed that scientific statements must be falsifiable. In Popper's view, only when knowledge is propositional, public, and falsifiable can both science and politics remain free, self-correcting, and genuinely progressive.

> *We think Popper would feel at home in a jurisdiction that celebrated propositional truth, thus near the Knowledge attractor, far away from Power, while Market and Morals remain perceptible. He experienced his time, we think, as dominated by chaos resulting from authoritarian clashes between various Power and morality claiming institutions.*

## Ian Goodfellow,

```
env,2,3,2,2
Pub,2,3,2,2
```

Ian Goodfellow (b. 1985)'s lived in the institutional landscape of tech capitalism, globalized research, and algorithmic governance in the early 21st century. This form of capitalism has a strong bias toward pragmatics over propositional truth—not because truth is rejected, but because functional performance, optimization, and emergent behavior have become the dominant criteria of success. Goodfellow entered this field during a time of massive investment in artificial intelligence, driven by cloud computing, big data, and deep learning breakthroughs. His work reflects the privatisation of cutting-edge research: after studying, he worked at Google Brain, OpenAI, and Apple—all at the frontier of AI development. The most powerful institutions of Goodfellow's career have not been governments or universities, but global tech corporations. These are hybrid institutions—nominally private but deeply entangled with state agendas, global markets, and military funding (e.g., through DARPA or dual-use AI concerns). The epistemic institutions surrounding Goodfellow are fast-moving, open-access, and pragmatically competitive (e.g., arXiv, NeurIPS, ICML), favoring results that demonstrate empirical improvement over theoretical finality.

Goodfellow's [29] work in machine learning—especially GANs—is part of a broader epistemic turn in AI away from truth as correspondence and toward functionality, emergence, and simulation. GANs work by adversarial training between a generator and a discriminator, with no explicit "ground-truth" supervision. They approximate realism without direct reference to propositional truth. This is a prime example of a system that produces plausible outputs through internal dynamics, rather than through correspondence to fixed external truths. This reflects a shift from understanding the world toward managing models that work well in it. Goodfellow's GANs became widely accepted not because they were "true," but because they produced images more realistic than any previous method. GANs (like LLMs, we claim) don't just mimic truth—they generate synthetic reali-ties. Goodfellow has been publicly concerned with AI safety, bias, and adversarial vulnerability—indicating that in a pragmatic epistemic (Market) regime, dealing with the consequences anchors in Morality and Power..

> *We think Goodfellow would feel at home in a jurisdiction where he was close to the Market attractor, where Power, Morality and Knowldege play second vikolin. We think his profile equals that of his Geo-political environment.*

### Summing up

What's at stake in this profiling exercise is the search for a shared jargon that can express core J4CC forces in each institutional zone without becoming intolerable in others. Our assumption is that reducing J4CC's jargon to 6 named dimensions (id,type,os,pc,hd,dp) will bring us a long way there. The goal is not epistemic unification but expressive scope and institutional interpretation—and thereby a workable path toward coordination. Thus far we are happy with the case study results, especially because it seamlessly supports profiling both individuals and communities. But how can we profile dynamics?

Lets consider another Case study.

## 2.2 Trump II – its first 167 days

The second case study is for further finding out about the frame's *prima facie* expressive qualities for profiling the dynamics in an institution's profile.

### Profile Dynamics

Consider a U.S. service provider (SP in this subsection) handling scientific data transfer within the U.S. on day 0 (zero) of the second Trump administration. Considering that weights are qualitative, we profile the political environment of the U.S. as we would have profiled the Biden administration: *low* Power weight (minimal state control), *high* Market weight (capitalist and commercial incentives dominate), *moderate* Morality weight (at least some ethical norms shaped by litigation, religion or lobbying are taken seriously) and *moderate* Knowledge weight (evidence based standards apply, but representation standards do less) or, in a vector, as [SP,USA,1,3,2,2]. Weights are derived from our qualifications of local regulatory density (Power), market share of private providers (Market), published ethics guidelines (Morality) and peer-review norms (Knowledge).

Now look at the political environment for our service provider on day 100 of "Trump II" (30 April 2025.) Some shifts suggest a re-balancing of informational authorities. Stock market movements are increasingly influenced by political signals, and governmental pressure has mounted on scientific and media institutions to align with its narratives. These developments reflect a relative strengthening of Sovereignty-like forces within the J4CC ecology of the U.S.—where state power, rather than procedural norms and scrutiny begins to shape information flows more directly. At day 100 of Trump II, our profiling vector of the U.S. has shifted from [SP,USA,1,3,2,2] to [SP,USA,2,3,2,1].

Now, to support reasoning about dynamics in J4CC frames we need to expand the 6-value list with a time stamp. When we do, the two completed vectors for service provider SP read:
[    day 1,SP,USA,1,3,2,2]
[day 100,SP,USA,2,3,2,1]

The trend visible here intensified by day 139 (7 June 2025), when President Trump federalized 4,000 California National Guard troops and ordered 700 Marines into Los Angeles under the Insurrection Act—over Governor Newsom's objections. The U.S. profile is, we think, now tending towards further reduction of the Moral attractor into:
[day 139,SP,USA,2,3,1,1].

Of course, data collection remains from now on required to trace the scope and durability of the U.S. framing shift noticed in our case study. Actually we continued monitoring until today, day 167. What we consider relevant is the U.S.-facilitated bombing attack on Iran of day 155 (without parliament being even notified, let alone consulted) and the signing of the OBBBA act today. Again we change the profile of the U.S. for expressing the grown Power attractor– it now becomes [day 167,SP,USA,3,3,1,1].

Reactions across mainstream and social platforms, though polarized, converged on the insights expressed in the vectors: the Power locus of informational control is being actively re-engineered in a direction that heightens its weight and keeps the Market attractor at its level of prominence while gradually reducing the weights of Morality and Knowledge.

### Summing up

First we show (in Table 1) how the frame allows to usefully profile the dynamics in the environment where an American service provider SP must find its feet during the first days of the second Trump administration:

*Table 3: Dynamic institutional vectors for Trump II*

| Time stamp | Agent | Env | os | mc | hd | dt |
|---|---|---|---|---|---|---|
| Day 0 | SP | USA | 1 | 3 | 2 | 2 |
| Day 100 | SP | USA | 2 | 3 | 2 | 1 |
| Day 139 | SP | USA | 2 | 3 | 1 | 1 |
| Day 167 | SP | USA | 3 | 3 | 1 | 1 |

Again, our results are recognized by Trump supporters ànd by those who wish him away: a trend towards enlarging the power attractor, toward consolidating Market and toward reduced influences of Morals and Knowledge.

We derive two insights from this case study. First, the frame is sufficiently expressive in natural language to articulate complex informational dynamics, even when dealing with imprecise values or partially obscured institutional structures. Second, we have by now some support for J4CC profiling rule-preparing (or political) agents' debates. What we did not yet look at is whether it is useful for expressing different and seemingly incommensurable political arguments.

Reason for yet another case study.

## 2.3 Constitutional Challenges

As we did not yet study a case that revealed if and how J4CC performs when confronted with the task to profile what formulations of debater X are intolerable to debater Y (and back again) in debating institution Z. We decided to take the dynamics that the first part of the Trump II administration realized as a debating subject and find the incompatible arguments in Klein-Taylor [37] and Bobbitt [38]. We feel they provide contrasting diagnoses of contemporary Power challenges through Klein-Taylor's "end times fascism" thesis versus Bobbitt's constitutional degradation analysis.

### Profiling the debate

The Klein-Taylor versus Bobbitt debate represents a paradigmatic case of cross-constitutional dialogue breakdown—two sophisticated intellectual positions that become mutually intolerable despite addressing identical phenomena. Using J4CC profiling, we can map how their force configurations create interpretive incompatibility.

*Klein-Taylor's Position Profile:* We code Klein-Taylor's "end times fascism" argument as:
[2025,Klein-Taylor,pub,1, 1, 3, 3].

Their analysis exhibits low Power weight (1)—they reject the legitimacy of current state power as "monstrous." Low Market weight (1) reflects their critique of oligarchic capitalism as complicit in civilizational collapse. High Morality weight (3) drives their revolutionary call for collective survival ethics. High Method weight (3) acknowledges their belief in scientifically framed propositional truth.

Their environmental context—academic/activist institutions during Trump II—profiles as:
[2025,Klein-Taylor,env, 3, 2, 2, 2].

High Power reflects the political pressure they experience from state power. Moderate Market (2) captures the mixed academic-commercial publication environment. Moderate Morality (2) represents the contested ethical landscape of contemporary academia. Moderate Method (2) reflects science institutions' move to displace part propositional truth priority with AI pragmatics, at least partly.

*Bobbitt's Position Profile:* Bobbitt's constitutional degradation analysis codes as:
[2025,Bobbitt, pub, 2, 2, 3, 2].

Moderate Power (2) reflects his acceptance of legitimate state authority while criticizing specific exercises of power. Moderate Market (2) acknowledges economic forces without making them central. High Morality (3) emphasizes rule of law and constitutional values, including prioritizing legal reasoning and precedent analysis, leaving propositional truth (Method) a moderate force.

His institutional environment—elite law schools and policy journals—profiles as:
[2025, Bobbitt, env, 2, 2, 3, 2].

This institutional ecology supports constitutional reasoning through established academic and legal channels, with moderate deference to both state authority and market forces while maintaining strong commitments to legal methodology and constitutional morality.

*Mapping Incompatibility:* The J4CC frame reveals why their arguments become mutually intolerable. Klein-Taylor's revolutionary language ("unhinged traitors," "monstrous survivalism") violates Bobbitt's high Morality commitment to measured legal discourse. Conversely, Bobbitt's "prudential approach" language violates Klein-Taylor's high Morality commitment to urgent collective action.

More fundamentally, their environmental force configurations create incompatible semantic expectations. Klein-Taylor operate within institutional contexts where apocalyptic language signals appropriate moral response to existential threats. Bobbitt operates within institutional contexts where measured constitutional analysis signals professional competence and democratic responsibility.

The force differential in Method proves particularly crucial. Klein-Taylor's high propositionally oriented Method weight (3) allows subordinating Power to moral urgency when addressing civilizational threats. Bobbitt's high Morality weight (3) requires maintaining moral analytical rigor even during political crises. This creates a semantic incompatibility: what Klein-Taylor see as appropriate moral response, Bobbitt sees as constitutional breakdown.

*Translation Possibilities:* Despite apparent incommensurability, J4CC suggests potential translation mechanisms. Klein-Taylor's "end times fascism" could be expressed in Bobbitt's jargon as "unprecedented constitutional crisis requiring institutional innovation." Bobbitt's "prudential approach" could be expressed in Klein-Taylor's jargon as "systematic legal resistance to authoritarian capture."

The frame reveals that both positions share high Morality commitments—they differ on institutional expression rather than fundamental values. Klein-Taylor's revolutionary language and Bobbitt's constitutional language both aim to preserve democratic possibility, but through incompatible institutional pathways.

**Summing up**

This case study demonstrates J4CC's capacity to profile seemingly incommensurable political arguments and identify both sources of incompatibility and potential translation pathways. The Klein-Taylor versus Bobbitt debate illustrates how identical empirical observations (Trump II's Power expansion) generate mutually intolerable interpretations when filtered through different institutional force configurations.

Three insights emerge. First, apparent ideological incommensurability often reflects institutional force differentials rather than fundamental value conflicts—both Klein-Taylor and Bobbitt exhibit high Morality commitments despite incompatible rhetoric. Second, environmental force configurations shape semantic tolerability as much as individual intellectual positions do—what counts as appropriate discourse varies dramatically across institutional contexts. Third, J4CC's vector notation enables systematic analysis of translation possibilities by identifying shared force commitments that could support cross-institutional dialogue.

The frame thus fulfills its promise of providing structured vocabulary for dialogue across incompatible institutional philosophies. Rather than eliminating disagreement, J4CC enables productive engagement by making explicit the underlying force weightings that drive constitutional conflicts. This suggests that our proposed jargon could indeed facilitate cross-constitutional dialogue during periods of institutional crisis.

# 3. Conclusion and Perspective

## 3.1 Conclusion

Our case studies lead us to the conclusion that J4CC in its current form already is a useful tool for analyzing political debate failures. Actually, we were surprised at how powerful the frame proved in practice, and how naturally it adapted with each case study. Yet looking at our problem statement

> *Can we meaningfully analyze domestic and cross-border debates about communication control and the accountability gap by developing a public jargon with a frame tailored to this purpose (J4CC)?*

we admit that our work thus far has provided a frame rather than a jargon (see Section 1.3.)

## 3.2 Perspective: Towards an AI-supported Model

So here we are, with a tentatively tested rational frame for a problem that does not allow nor even attempt Knowledge's blessing. Its Knowledge-related scrutiny has been augmented, some will say contaminated, with Power-related security, Market-related pragmatism and Morality-related virtue. Intelligent human agents seem to be able to handle this complexity when considering a case. To us, employed in humanities oriented departments, it does not feel too outlandish to consider the possibility of designing an AI language model to help us here. After all, a jargon has a close family resemblance to a language. In the remaining subsection we theorize at the everyday language level we command about a few hurdles that have to be taken.

*Why an AI language model?*

But why would we do that? Why would we attempt to implement the frame into a successful language model? Isn't J4CC in the current form already a useful debate-supporting tool? We think in practice it is already. But we also feel that how vectors are coded around debate topics, debating agents and their environments is still too speculative to gain academic blessing. Part of the reason is that our frame can only gain traction by proving its pragmatic usefulness, just like language models as ChatGPT and Claude do. Language models are useful, their operation is neither true to nor semantically understood by the agents that make or use them. Still they gained so much usefulness in a few years that by now they are by default augmenting browsers and search engines, which are ubiquitous, supporting at least billions of users who rely on these tools daily without understanding their internal mechanisms—precisely the kind of pragmatic adoption J4CC must achieve to research cross-constitutional dialogues on a grand scale. J4CC's jargon is a language, and modern AI trains and grows general language models, and connects them with specialist knowledge bases, even feeds them. How is our jargon different from a specialist knowledge base? The answer lies in Anderson's [27] insights about emergence (see Section 1.5)

This suggests that J4CC's jargon cannot be simply programmed but must emerge from complex interactions between the frame, training data, and use patterns—precisely what large language models excel at generating. Our current work is essential input for the functional design needed for realizing a dedicated RAG-LLM pipe.[11] For our sequel, our eyes are already looking at what Anthropic's interpretive projects are doing (e.g., Bai et al., 2022 and Templeton, A. et al., 2024). Whether such a sequel project is feasible is outside the scope of this article – we moved it to the appendix.

Whether J4CC can achieve this emergent transformation from frame to jargon remains an open empirical question. But the convergence of constitutional crisis, institutional breakdown, and AI capability suggests that the attempt is both necessary and timely. The alternative—continued dialogue failure across institutional divides—offers no path forward in an age of global governance challenges.

---

[11] Again: a few technical terms that have not yet entered everyday language.

# Appendix – Is Implementation of J4CC Feasible?

A follow-up implementation project could be structured through a contract with clearly defined rules. We assume that a successful stakeholder debate has already occurred—at least among the authors (AS and KZ), who remain motivated to proceed.

Given that conducting a comprehensive stakeholder debate is beyond our present scope, we propose a hypothetical pilot study aimed at generating interest from academia and potential investors. This involves outlining a phased approach to constructing a Retrieval-Augmented Generation (RAG)-style pipeline for J4CC, along with an initial mapping of associated risks and opportunities. To estimate feasibility and minimum required investment, we prompted an LLM as follows:

> "Imagine we would like to use Wikipedia as an external vector base for an implementation of J4CC as an open-source seq2seq model. What would the minimum investment be?"

Below is the proposed roadmap based on the response:

**Phases**

1. Data Collection and Preparation
   - Gathering transcripts or minutes from public debates, ensuring content feasibility and legal usability.
2. Data Encoding into Vector Embeddings
   - Utilizing BAAI's `gpe-large` model. The software is open-source, but the hardware requirement includes at least one NVIDIA A100 GPU.
3. Vector Database and Retrieval Layer Setup
   - Deploying Pinecone for vector storage and retrieval, an affordable solution suitable for prototyping.
4. Generator Integration (seq2seq model)
   - Selecting an appropriate Large Language Model (LLM) as a generator. Options include:
     - **Open-source models:** T5, Mistral, or Falcon, deployed locally.
     - **Commercial API-based models:** ChatGPT (OpenAI), Claude (Anthropic), or DeepSeek.

Each of these options presents distinct trade-offs regarding control, transparency, cost, and geopolitical considerations—critical factors given J4CC's emphasis on zone-specific filters and ethical guidelines.

## Production Cycle
5. User Query → Retriever → Prompt Construction → Generator → Generated Answer

## J4CC Risks and Constraints
Each phase faces unique J4CC-specific challenges:
- **Pow**er: Can we ethically and legally use tools like gpe-large developed in Beijing? While open-source solutions enable auditing and self-hosting, commercial API solutions such as ChatGPT or Claude entail data flow to U.S.-based servers, introducing geopolitical and regulatory complexities.
- **Market:** Are the GPU compute and hosting requirements affordable? What viable options exist for open hardware access? API-based models typically charge per token, becoming costly at scale. Open-source models incur upfront GPU expenses but have negligible ongoing costs once deployed.
- **Morality:** What types of content must be filtered or safeguarded within the retrieved corpus? Are specific answers unacceptable or sensitive? API-based models embed opaque filtering mechanisms without transparency. In contrast, open-source models allow explicit definition and control of ethical boundaries.
- **Knowledge:** How transparent is the generator regarding its training datasets, fine-tuning procedures, and internal guardrails? API-based providers do not disclose detailed training data or fine-tuning methods, while open-source models offer full inspection and fine-tuning flexibility.

## Assessmemt
Given that the LLM's responses to our prompt exhibited no attempts at confabulation and could be independently verified, we consider that implementing J4CC within a RAG-LLM pipeline architecture is indeed feasible.